\documentclass[aps,prd,twocolumn,nofootinbib]{revtex4-1}
\usepackage{epsfig}
\usepackage[colorlinks,linkcolor=blue,anchorcolor=blue,citecolor=blue,urlcolor=blue,breaklinks=true]{hyperref}
\usepackage{graphics}
\usepackage{color}
\usepackage{slashed}
\usepackage{makecell}
\usepackage{subfigure}
\usepackage{changes}
\usepackage{cancel}

\begin{document}
\author{Shu-Sheng Xu$^{1}$}~\email[]{Email: xuss@njupt.edu.cn}
\address{$^{1}$ School of Science, Nanjing University of Posts and Telecommunications, Nanjing 210023, China}

\title{Pion superfluid phase transition at finite isospin chemical potential}
\begin{abstract}
The pion superfluidity phase transition at $T-\mu_I$ and $\mu_q-\mu_I$ planes are studied in the framework of Dyson-Schwinger equations. The rainbow truncation and Gaussian effective gluon propagator are employed to calculate pion condensate, $\langle\pi\rangle$, as a function of $\mu_I$ at finite $T$ and $\mu_q$. At $T=(0, 80, 113, 120)$~MeV, the $\langle\pi\rangle$ keeps zero when $\mu_I$ is less than a critical value $\mu_{Ic}(T)$. The $\langle\pi\rangle$ at $T=113$~MeV agrees with the lattice QCD result. At the $T-\mu_I$ plane, the pion superfluid phase appears at low temperature and high isospin chemical potential region. At finite $\mu_q$, $\langle\pi\rangle$ varying with $\mu_I$ almost coincide for $\mu_q<400$~MeV. At $\mu_q-\mu_I$ plane, the pion superfluid phase appears at $\mu_I\gtrsim m_\pi$ when $\mu_q<400$~MeV.

\bigskip


\bigskip

\noindent PACS Number(s): 11.30.Rd, 25.75.Nq, 12.38.Mh, 12.39.-x

\end{abstract}
\maketitle

\section{Introduction}\label{intro}
Hot and~(or) dense quantum chromodynamics (QCD) systems play an essential role in many issues. Hot strong interacting quarks and gluons, namely quark-gluon plasma~(QGP), are present in the evolution of the early universe, the phase transition from QGP to hadrons was presumed during the process of cooling~\cite{annurev.nucl}. Relativistic heavy-ion collisions at facilities, such as Relativistic Heavy Ion Collider (RHIC) and CERN heavy-ion program, can produce such high-temperature~($T$). Neutron (or quark) stars are the unique natural dense QCD systems.

Thermal QCD matter is of particular interest since the lattice simulations can be performed, which is the only way to solve the QCD non-perturbatively from the first principle to the desired precision.
However, there is an infamous difficulty, that is the ``sign problem'', in studying the properties of dense QCD systems. Various effective models are proposed to study QCD systems at finite baryon chemical potential $\mu_B$, such as chiral quark model~\cite{PhysRevC.69.065802,JAKOVAC2004179}, Nambu-Jona-Lasinio (NJL) model~\cite{BUBALLA2005205,Fukushima_2010,cui2014wigner,PhysRevD.94.014026,PhysRevD.82.056006,Andersen_2009}, functional renormalization group~\cite{schaefer2008renormalization,andersen2012chiral,DREWS201769}, the sum rules~\cite{PhysRevD.42.1744,MANUEL2001200,ayala2017finite,PhysRevD.92.016006}, and some specific models in the framework of Dyson-Schwinger equations~(DSEs)~\cite{PhysRevD.94.094030,PhysRevD.82.034006,PhysRevLett.106.172301,PhysRevD.84.014017,fischer2019qcd,PhysRevD.100.074011,PhysRevD.90.034022,jiang2013chiral,shi2014locate}. Based on these studies, people commonly believed that the strong interaction matter suffers a phase transition from the hadronic matter at low temperature and density to quark-gluon plasma~(QGP) at high temperature and density.

At the extremely high baryon chemical potential, the dense QCD system is weakly interacting since asymptotic freedom, the dominant interaction between quarks are one gluon exchange, which gives rise to color superconductivity~\cite{PhysRevD.61.074017,PhysRevD.61.051501,RevModPhys.80.1455}.
Based on the above-mentioned model studies, people believe that there are rich structures in the region of finite baryon chemical potential together with finite isospin chemical potential, for instance, flavor-color locked~(FCL) phase~\cite{PhysRevLett.95.152002,PhysRevD.66.074017,PhysRevD.76.105030}, pion and kaon superfluid phases~\cite{PhysRevLett.86.592,PhysRevD.89.116006,particles2030025,PhysRevD.74.036005,PhysRevC.75.064901}.

At the finite $\mu_I$ system with $T=0$ and $\mu_B=0$, chiral perturbation theory and lattice regularized QCD~(LQCD) have claimed that there is a critical isospin chemical potential $\mu_I^c=m_\pi$, after this point the pion superfluid phase appears~\cite{PhysRevLett.86.592,adhikari2019two,KOGUT2004542}. It is also confirmed by many model studies~\cite{PhysRevD.79.034032,sym11060778,PhysRevD.71.116001}. And many other studies on the case of finite temperature~\cite{TOUBLAN2003212,lu2020thermodynamics,BARDUCCI2003217}, which reveal that a pion superfluid phase transition still appears.
At present, a study on the pion condensate and pion superfluid phase transition in the framework of DSEs is still absent.

The equation of state~(EoS) at finite $\mu_B$ and $\mu_I$ plays key role to the structures of neutron stars and possible quark stars, which is strongly related to the phase structures of dense QCD system. Nowadays, the structures of neutron stars are studied extensively in the quasi-particle model and NJL model~\cite{PhysRevD.86.114028,PhysRevD.92.054012,zhao2017phenomenological,PhysRevD.95.056018,xu2021qcd}. Neutron stars with DSEs approach also have studied~\cite{BaiQCD}. People guess that the pion condensate may appear in the core of neutron stars. Recently, pion stars, QCD matter in pion superfluid phase, are discussed primarily~\cite{PhysRevD.98.094510}.

In this paper, we will study the phase structures at finite $\mu_I$, specifically on $T-\mu_I$ and $\mu_q-\mu_I$ plane, in the framework of DSEs. The pion condensate is studied in detail, which is used to determine the pion superfluid phase transitions. This paper is organized as follows. In Sec.~\ref{dsesandgp}, we give an introduction to the DSEs at finite temperature and finite chemical potentials, the employed truncation and approximations are clarified. The pion condensate varying with $\mu_I$ at some specific values of temperature are shown in Sec.~\ref{results}, which help us to determine the location of pion superfluid phase transition. Thereafter, the pion superfluid phase diagram on the $T-\mu_I$ plane is displayed and compared with LQCD. Afterward, the pion condensate and pion superfluid phase transition at finite $\mu_q$ and $\mu_I$ are discussed. We give a summary in Sec.~\ref{sum}.

\section{Quark Dyson-Schwinger equation at finite temperature and chemical potentials}\label{dsesandgp}
The Dyson-Schwinger equations are general relations of Green functions of quantum field theory. The formula of quark DSE at finite $\mu_I$ is different with the case of $\mu_I=0$ since the vacuum expectation value of $\bar q \gamma_5\vec{\tau} q$ unequal to zero, $\langle\Omega| \bar q \gamma_5\vec{\tau} q |\Omega\rangle\neq 0$, in some region of phase diagram. Therefore, the quark DSE should be rederived from QCD generating functional, and the quark DSE is given by,
\begin{eqnarray}
S^{-1}(\vec{p},\tilde{\omega}_p) &=& Z_2 S_0^{-1}(\vec{p},\tilde{\omega}_p) + \Sigma(\vec{p},\tilde{\omega}_p),	\label{quarkDSE}
\\
\Sigma(\vec{p},\tilde{\omega}_n) &=& \frac{16\pi}{3}\frac{Z_2}{\tilde{Z}_3} \sum_{n_q}\!\!\!\!\!\!\!\!\int\mathrm{Tr}\Big[ g^2D_{\mu\nu}(\vec{k},\Omega_k)
\nonumber\\
&&\times \gamma_\mu S(\vec{q},\omega_q) \Gamma_\nu(\vec{p},\tilde{\omega}_p,\vec{q},\tilde{\omega}_q) \Big],
\end{eqnarray}
where $S^{-1}(\vec{p},\tilde{\omega}_p)$ is the inverse of quark propagator with two flavors, with $\tilde{\omega}_p = \omega_p + i\mu_f$ and the Matsubara frequency $\omega_p=(2n_p+1)\pi T$. $S_0^{-1}(\vec{p},\tilde{\omega}_p)$ is the inverse of bare quark propagator, $\Sigma(\vec{p},\tilde{\omega}_p)$ is the quark self-energy. $g$ is strong coupling constant and $D_{\mu\nu}(\vec{k},\Omega_k)$ is dressed gluon propagator with $\vec{k}=\vec{p}-\vec{q}$ and $\Omega_k=\tilde\omega_p-\tilde\omega_q$. The $\Gamma_\nu(p,q)$ is the dressed quark-gluon vertex. $Z_2$, $\tilde{Z}_3$ are the quark wave function and ghost renormalization constants respectively. The brevity
\begin{equation}
\sum_{n_q}\!\!\!\!\!\!\!\!\int = T\sum_{n_q} \int\frac{d^3\vec{q}}{(2\pi)^3}.
\end{equation}
The quark propagator have form
\begin{eqnarray}
S(\vec{p},\tilde\omega_p)=
\left(
\begin{array}{cc}
S^{uu}(\vec{p},\tilde\omega_p) 	&S^{ud}(\vec{p},\tilde\omega_p)	\\
S^{du}(\vec{p},\tilde\omega_p) 	&S^{dd}(\vec{p},\tilde\omega_p)	
\end{array}
\right),
\label{quarkprop}
\end{eqnarray}
with
\begin{eqnarray}
S^{aa}(\vec{p},\tilde\omega_p) &=& i\slashed{\vec{p}} \sigma^a_A(\vec{p},\tilde\omega_p) + \sigma^a_B(\vec{p},\tilde\omega_p) 
\nonumber\\
&&+ i\gamma_4\omega_p \sigma^a_C(\vec{p},\tilde\omega_p) + \slashed{\vec{p}} \gamma_4 \omega_p \sigma^a_D(\vec{p},\tilde\omega_p),	\label{quarkprop1}
\\
S^{ab}(\vec{p},\tilde\omega_p) &=& i\gamma_5\Big(i\slashed{\vec{p}} \sigma^{ab}_{A_5}(\vec{p},\tilde\omega_p) + \sigma^{ab}_{B_5}(\vec{p},\tilde\omega_p)
\nonumber\\
&&+ i\gamma_4\omega_p \sigma^{ab}_{C_5}(\vec{p},\tilde\omega_p) + \slashed{\vec{p}} \gamma_4 \omega_p \sigma^{ab}_{D_5}(\vec{p},\tilde\omega_p)\Big).	\label{quarkprop2}
\end{eqnarray}
The notation $aa$ can be $aa=uu,dd$ and $ab$ can be $ab=ud,du$.
It is widely accepted that the two-point Green function with one u quark external leg and one d quark external leg is vanished at low temperature and density, while it would be finite in the pion superfluid phase.  Hence, the quark propagator with different flavor, namely $S^{ud}(\vec{p},\tilde{\omega}_p)$ and $S^{du}(\vec{p},\tilde{\omega}_p)$, survives. In this case, the quark DSE is formally the same as the case of $\mu_I=0$, but the quark propagator has the form of Eq.~(\ref{quarkprop}). The inverse of quark propagator can be written as
\begin{eqnarray}
S^{-1}(\vec{p},\tilde\omega_p)=
\left(
\begin{array}{cc}
{S^{-1}}^{uu}(\vec{p},\tilde\omega_p) 	&{S^{-1}}^{ud}(\vec{p},\tilde\omega_p)	\\
{S^{-1}}^{du}(\vec{p},\tilde\omega_p) 	&{S^{-1}}^{dd}(\vec{p},\tilde\omega_p)	
\end{array}
\right),
\label{inversequarkprop}
\end{eqnarray}
with
\begin{eqnarray}
{S^{-1}}^{aa}(\vec{p},\tilde\omega_p)&=& i\slashed{\vec{p}} A^a(\vec{p},\tilde\omega_p) + B^a(\vec{p},\tilde\omega_p)
\nonumber\\
&&+ i\gamma_4\omega_p C^a(\vec{p},\tilde\omega_p) + \slashed{\vec{p}} \gamma_4 \omega_p D^a(\vec{p},\tilde\omega_p),	\label{inversequarkprop1}
\\
{S^{-1}}^{ab}(\vec{p},\tilde\omega_p) &=& i\gamma_5\Big(i\slashed{\vec{p}} A^{ab}_5(\vec{p},\tilde\omega_p) + B^{ab}_5(\vec{p},\tilde\omega_p)
\nonumber\\
&&+ i\gamma_4\omega_p C^{ab}_5(\vec{p},\tilde\omega_p) + \slashed{\vec{p}} \gamma_4 \omega_p D^{ab}_5(\vec{p},\tilde\omega_p)\Big).	\label{inversequarkprop2}
\end{eqnarray}
The relation between the eight scalar functions in Eqs.~(\ref{quarkprop1}-\ref{quarkprop2}) and the eight scalar functions in Eqs.~(\ref{inversequarkprop1}-\ref{inversequarkprop2}) is tedious and not shown here.
The dressed quark propagator depends on the dressed gluon propagator and the dressed quark-gluon vertex, they satisfy their own DSEs, which are infinitely coupled equations. Some specific truncation and approximations are inevitable, we employ rainbow truncation in this work, that is
\begin{equation}
g^2 D_{\mu\nu}(\vec{k},\Omega_k) \Gamma_\nu(\vec{p},\tilde{\omega}_p,\vec{q},\tilde{\omega}_q) \rightarrow \mathcal{G}(s) D^0_{\mu\nu}(\vec{k},\Omega_k) \gamma_\nu,
\end{equation}
with effective interaction model
\begin{equation}
\mathcal{G}(s) = d\frac{4\pi^2}{\sigma^6} s e^{-\frac{s}{\sigma^2}},    \label{gluonmodel}
\end{equation}
where $s=\vec{k}^2+\Omega_k^2$, and $D^0_{\mu\nu}(\vec{k},\Omega_k)$ is the free gluon propagator. Many studies have shown that the parameters, $d$ and $\sigma$, are not independent. For $\sigma\in [0.3, 0.5]~\mathrm{GeV}$, the masses and decay constants of ground $\pi$ and $\rho$ mesons are roughly constants if $d\sigma=(0.8~\mathrm{GeV})^3$.

It is worth noting that the effective gluon propagator $\mathcal{G}(s)$ is heavily suppressed in the ultraviolet region. The renormalization constants $Z_2$ and $\tilde Z_3$ are both set to one in this work. Substituting Eqs.~(\ref{quarkprop1}-\ref{gluonmodel}) into Eq.~(\ref{quarkDSE}), one can obtain the 8 coupled nonlinear integral equations.
\section{Pion superfluidity phase transition}\label{results}
In this section, we mainly focus on the phase structures at $T-\mu_I$ and $\mu_q-\mu_I$ planes with the help of pion condensate.
Pion condensate is regarded as the order parameter for the pion superfluid phase transition, it is defined as
\begin{eqnarray}
\langle\pi\rangle &=& \langle \bar u i\gamma_5 d\rangle + \langle \bar d i\gamma_5 u\rangle
\nonumber\\
&=&-N_c \sum_p \!\!\!\!\!\!\!\!\int \mathrm{tr}\left[ i\gamma_5\left(S^{ud}(\vec{p},\tilde\omega_p) + S^{du}(\vec{p},\tilde\omega_p) \right) \right].
\end{eqnarray}
\subsection{The relation between $m_\pi$ and the critical $\mu_I$}
In the framework of DSEs, the pion mass is the solution of the homogeneous Bethe-Salpeter equation~(BSE),
\begin{equation}
\Gamma_\pi(p,P) = \int \frac{d^4q}{(2\pi)^4} K(p,q,P) S(q_+) \Gamma_\pi(q,P) S(q_-),    \label{pionBSE}
\end{equation}
where $P$ is the meson momentum and $P=(0,0,0,im_\pi)$ in the meson rest frame. $\Gamma_\pi(p,P)$ is the Bethe-Salpeter amplitude of pion, $K(p,q,P)$ is the interaction kernel. $q_+=q+P/2$, $q_-=q-P/2$. In the rainbow-ladder truncation,
\begin{equation}
K(p,q,P) = -\mathcal{G}((p-q)^2) T_{\mu\nu}(p-q) \gamma_\mu \frac{\lambda^a}{2} \otimes \gamma_\nu \frac{\lambda^a}{2},  \label{kernel}
\end{equation}
with
\begin{equation}
T_{\mu\nu}(p-q) = \delta_{\mu\nu} - \frac{(p-q)_\mu (p-q)_\nu}{(p-q)^2},
\end{equation}
and $\mathcal{G}((p-q)^2)$ has been defined in Eq.~(\ref{gluonmodel}), $\lambda^a (a=1,2,\cdots,8)$ are Gell-Mann matrices. 
Inserting Eq.~(\ref{kernel}) into Eq.~(\ref{pionBSE}), the pion BSE can be written as
\begin{eqnarray}
\Gamma_\pi(p,P) &=& -\frac{4}{3}\int \frac{d^4q}{(2\pi)^4} \mathcal{G}((p-q)^2) T_{\mu\nu}(p-q) 
\nonumber\\
&&\hspace*{10mm}\times\gamma_\mu S(q_+) \Gamma_\pi(q,P) S(q_-) \gamma_\nu,    \label{pionBSE}
\end{eqnarray}
In general, the $\Gamma_\pi(p,P)$ can be decomposed as
\begin{eqnarray}
\Gamma_\pi(p,P) &=& i\gamma_5\Big(E_\pi(p,P) + i\slashed{P} F_\pi(p,P)
\nonumber\\
&&\hspace*{5mm} + i \slashed{p} G_\pi(p,P) + \slashed{P}\slashed{p} H_\pi(p,P)\Big).
\end{eqnarray}
It is equivalent to the structure of $S^{ud}$ in Eq.~(\ref{quarkprop2}) if $P=(0,0,0,im_\pi)$ are taken into account. It is worth noting that the argument of quark propagator $q_\pm^2=(q\pm P/2)^2$ is equivalent to take the replacement $q_4\rightarrow q_4 \pm im_\pi/2$ in the propagator.

We now turn attention to the two flavor quark propagator at finite $\mu_I$, it satisfies
\begin{eqnarray}
S^{-1}(p,\mu_I) &=& S_0^{-1}(p,\mu_I) + \Sigma(p,\mu_I),    \label{quarkprop3}
\\
\Sigma(p,\mu_I) &=& \frac{4}{3}\int \frac{d^4q}{(2\pi)^4} \mathcal{G}((p-q)^2) T_{\mu\nu}(p-q) \gamma_\mu S(q,\mu_I) \gamma_\nu,    \label{quarkprop4}
\nonumber\\
\end{eqnarray}
at $T=0$ and $\mu_q=0$.

Assuming the mixed flavor parts of quark propagator, namely $S^{ud}$ and $S^{du}$, are greatly smaller than $S^{uu}$ and $S^{dd}$. It is apparently true around critical $\mu_I$, because $S^{ud}=0=S^{du}$ in this region. The quark propagator can be expended as
\begin{eqnarray}
S(p,\mu_I) = S_0(p,\mu_I) + S_1(p,\mu_I),   \label{quarkprop5}
\end{eqnarray}
where $S_0(p,\mu_I)=\mathrm{diag}(S_0^{uu}(p_+),S_0^{dd}(p_-))$ with $p_\pm=(\vec{p},p_4\pm i\mu_I/2)$, which is the solution with conditions of $S^{ud}=S^{du}=0$, namely $S^{uu}(p_+)$ and $S^{dd}(p_-)$ satisfy their own DSEs.

Similarly, the inverse of quark propagator have 
\begin{equation}
S^{-1}(p,\mu_I) = S_0^{-1}(p,\mu_I) + S_1^{-1}(p,\mu_I).    \label{quarkprop6}
\end{equation}
Making use of $SS^{-1}=S^{-1}S=1=S_0S_0^{-1}=S_0^{-1}S_0$, one have
\begin{eqnarray}
S_0(p,\mu_I) S_1^{-1}(p,\mu_I) &=& - S_1(p,\mu_I) S_0^{-1}(p,\mu_I),    \label{cond1}
\\
S_0^{-1}(p,\mu_I) S_1(p,\mu_I) &=& - S_1^{-1}(p,\mu_I) S_0(p,\mu_I).    \label{cond2}
\end{eqnarray}
The quark DSEs, Eqs.~(\ref{quarkprop3}) and (\ref{quarkprop4}), have two parts, diagonal and non-diagonal. The diagonal parts are simply the $u$ and $d$ quarks' DSEs at $\langle\pi\rangle=0$, while $u-d$ mixed part is
\begin{eqnarray}
S^{-1ud}(p,\mu_I) = \frac{4}{3}\int \frac{d^4q}{(2\pi)^4} \mathcal{G}((p-q)^2) T_{\mu\nu}(p-q) \gamma_\mu S^{ud}(q,\mu_I) \gamma_\nu, \label{quarkDSE2}
\end{eqnarray}
where $S^{ud}(q,\mu_I)$ can be written as
\begin{eqnarray}
S^{ud} &=& \left( S S^{-1} S \right)^{ud}
\nonumber\\
&=& \left( (S_0 +S_1) (S_0^{-1} +S_1^{-1}) (S_0+S_1) \right)^{ud}
\nonumber\\
&=& \left( S_1 S_0^{-1} S_0 + S_0 S_1^{-1} S_0 + S_0 S_0^{-1} S_1 \right)^{ud}
\nonumber\\
&=& - \left( S_0 S_1^{-1} S_0 \right)^{ud}.
\end{eqnarray}
In the last step, the Eqs.~(\ref{cond1}) and (\ref{cond2}) have been used. The diagonal parts are all omitted, because we concentrate the non-diagonal part. In order to show the concrete arguments of quark propagators, the details of $S^{ud}$ is
\begin{eqnarray}
&&\left(
\begin{array}{cc}
0   &S^{ud}(q,\mu_I)  \\
S^{du}(q,\mu_I)   &0
\end{array}
\right)
\nonumber\\
&=&-
\left(
\begin{array}{cc}
S_0^{uu}(q_+)   &0  \\
0   &S_0^{dd}(q_-)
\end{array}
\right)
\left(
\begin{array}{cc}
0   &S_1^{-1ud}  \\
S_1^{-1du}   &0
\end{array}
\right)
\nonumber\\
&&\hspace*{2mm}\times
\left(
\begin{array}{cc}
S_0^{uu}(q_+)   &0  \\
0   &S_0^{dd}(q_-)
\end{array}
\right)
\nonumber\\
&=&-
\left(
\begin{array}{cc}
0   &S_0^{uu}(q_+)S_1^{-1ud}S_0^{dd}(q_-)  \\
S_0^{dd}(q_-)S_1^{-1du}S_0^{uu}(q_+)   &0
\end{array}
\right).\label{quarkprop7}
\end{eqnarray}
The non-diagonal quark propagator turns to
\begin{eqnarray}
S^{-1ud}(p,\mu_I) &= &-\frac{4}{3}\int \frac{d^4q}{(2\pi)^4} \mathcal{G}((p-q)^2) T_{\mu\nu}(p-q)
\nonumber\\
&&\hspace*{1mm}\times \gamma_\mu S_0^{uu}(q_+) S^{-1ud}(q,\mu_I) S_0^{dd}(q_-) \gamma_\nu. \label{quarkDSE3}
\end{eqnarray}
Comparing Eqs.~(\ref{pionBSE}) and (\ref{quarkDSE3}), one finds that $S^{-1ud}(p,\mu_I)$ satisfy the same equation with $\Gamma_\pi(p,P)$. According to the experience of pion BSE, the $\gamma_5$ term is dominate in the four terms, which is proportional to the order parameter of pion superfluid phase transition, $\langle\pi\rangle$. Therefore, we prove that the pion superfluid phase transition happens at $\mu_I=m_\pi$ when $T=0$ and $\mu_q=0$.
\subsection{numerical results}
\begin{figure}[t]
\centering
\includegraphics[width=0.45\textwidth]{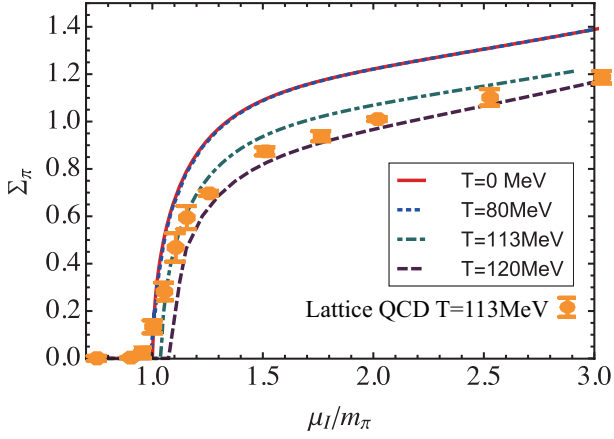}
\caption{The renormalized pion condensates varying with $\mu_I/m_\pi$ at four different temperatures. Comparing with LQCD in Ref.~\cite{PhysRevD.97.054514}.}
\label{fig1}
\end{figure}
\begin{figure}[b]
\centering
\includegraphics[width=0.45\textwidth]{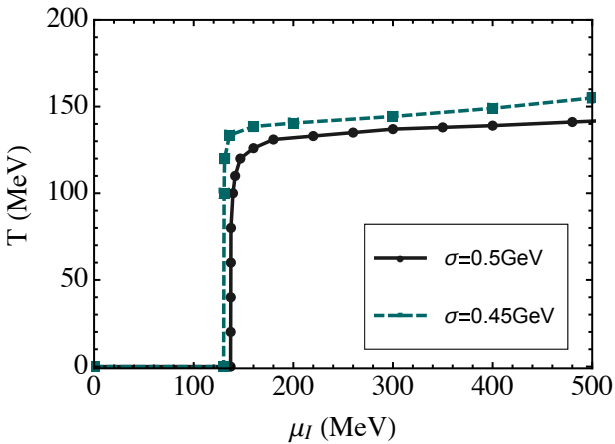}
\caption{The pion superfluid phase transition line in $T-\mu_I$ plane.}
\label{fig2}
\end{figure}

In order to show the numerical results, related parameters should be fixed in advance. In this work, we use $m_u=m_d=m_q=5$~MeV, $\sigma=0.5$~GeV and $d=1~\mathrm{GeV}^2$, which are fitted by pion mass and decay constant~\cite{PhysRevD.65.094026}. The renormalized pion condensate $\Sigma_\pi$, which is defined as~\cite{PhysRevD.97.054514}
\begin{equation}
\Sigma_\pi = \frac{m_q}{m_\pi^2 f_\pi^2} \langle\pi\rangle,
\end{equation}
varying with $\mu_I/m_\pi$ with four different $T$ are displayed in Fig.~\ref{fig1}. We can see from Fig.~\ref{fig1} that the renormalized pion condensate keeps vanish at $\mu_I/m_\pi<1$ at $T=0$, which implies the pion superfluid phase transition happens at $\mu_I^c\sim m_\pi$, which has been predicted by chiral perturbation theory and confirmed by LQCD. The curves of $T=80~\mathrm{MeV}$ is almost overlapped with the curves of $T=0$, which implies the critical $\mu_I$ is almost unchanged in the region of $T<80~\mathrm{MeV}$. Concerning on the $T=113~\mathrm{MeV}$ and $T=120~\mathrm{MeV}$, we can see that the critical $\mu_I$ becomes larger and larger, which indicates that the $\mu_I^c(T)$ increase progressively. We can see that the curve of pion condensate at $T=113$~MeV is agreement with the lattice data~\cite{PhysRevD.97.054514}. 
\begin{figure}[t]
\centering
\includegraphics[width=0.46\textwidth]{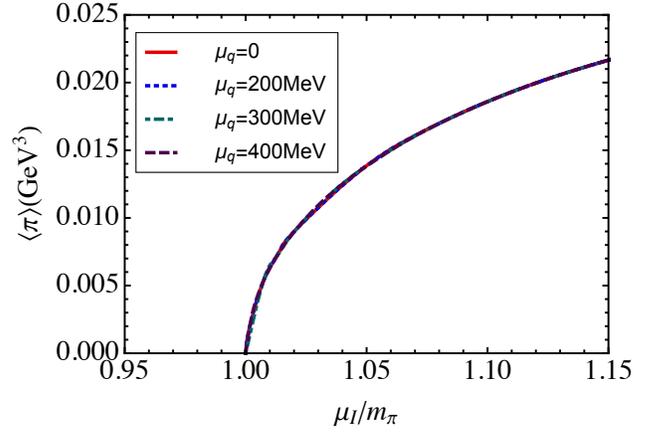}
\caption{The pion condensate varying with $\mu_I/m_\pi$ with quark chemical potentials.}
\label{fig3}
\end{figure}
\begin{figure}[b]
\centering
\includegraphics[width=0.46\textwidth]{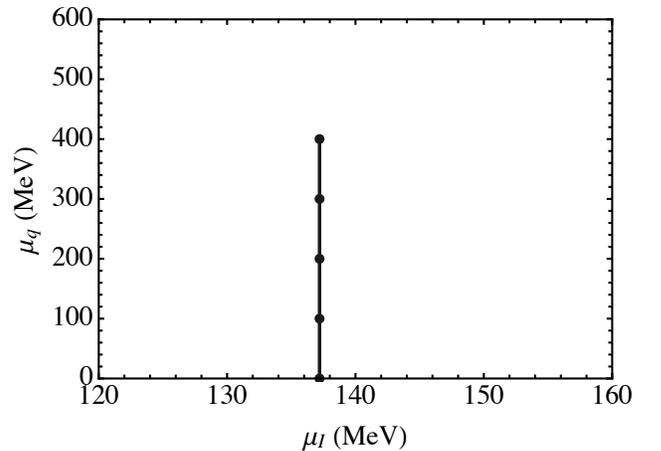}
\caption{The pion superfluid phase transition line in $\mu_q-\mu_I$ plane.}
\label{fig4}
\end{figure}

The pion superfluid phase diagram on $T-\mu_I$ plane can be plotted by connect $\mu_{Ic}$ for temperatures, which is illustrated in Fig.~\ref{fig2}. We can see in Fig.~\ref{fig2} that the critical $\mu_I$ almost unchanged in the region of $T\lesssim 0.1~$GeV. At relative high $\mu_I$, the critical temperature increases very slow and almost keeps a constant. The phase diagram with $\sigma=0.45~$GeV, $d=1.025~\mathrm{GeV}^2$~\cite{PhysRevD.65.094026} are also displayed to give a comparison, it shows that the phase transition line locates qualitatively same position when $\sigma$ decreases 10\%. To sum up, the pion condensate is nonzero at high isospin chemical potential and low temperature.

The phase structures at $\mu_q-\mu_I$ plane are crucial for the compact stars. In Fig.~\ref{fig3}, the pion condensate varying with $\mu_I$ at $\mu_q=(0, 200, 300, 400)$~MeV are displayed. All these curves are almost coincide, which implies the pion superfluidity phase transition locates $\mu_I\sim m_\pi$ for $\mu_q\in (0,400)$~MeV. The pion superfluid phase diagram at $\mu_q-\mu_I$ is shown in Fig.~\ref{fig4}. We can not draw the critical line with $\mu_q>400$~MeV because the quark DSEs are not converge numerically in this region. Such difficulty is also appeared in solving complex plane quark DSEs.


\section{Summary}\label{sum}
At finite isospin chemical potential, non-zero u-d and d-u quark propagators are assumed to derive the quark Dyson-Schwinger equations with rainbow truncation. The Gaussian effective gluon propagator is used to take concrete numerical calculations, where parameters are fixed by pion mass and its decay constant. We studied the pion superfluid phase transition at $T-\mu_I$ and $\mu_q-\mu_I$ plane with the help of pion condensate, $\langle\pi\rangle$. The $\langle\pi\rangle$ varying with $\mu_I$ at $T=(0, 80, 113, 120)$~MeV are plotted, the renormalized pion condensate as function of $\mu_I$ at $T=113$~MeV is agree with that of lattice QCD. Collecting the pion superfluid phase transition points, we plot the phase diagram at the $T-\mu_I$ plane. It shows that the critical $\mu_I$ is almost constant at $T<100$~MeV, and the critical temperature is almost unchanged at high $\mu_I$. The pion superfluid phase appears in the region of high $\mu_I$ and low $T$.

At $\mu_q-\mu_I$ plane, the pion condensate keeps zero at $\mu_I\lesssim m_\pi$ for $\mu_q<400$~MeV. All the curves of pion condensate as a function of $\mu_I$ almost coincide. It indicates that the critical $\mu_I$ keeps constant for $\mu_q<400$~MeV. In the region of $\mu_q>400$~MeV, the pion condensate and pion superfluid phase transition require further study.

\acknowledgments
This work is supported in part by the National Natural Science Foundation of China (under Grant No. 11905107), the National Natural Science Foundation of Jiangsu Province of China (under Grant No. BK20190721), Natural Science Foundation of the Jiangsu Higher Education Institutions of China (under Grant No. 19KJB140016), Nanjing University of Posts and Telecommunications Science Foundation (under grant No. NY129032), Innovation Program of Jiangsu Province.
\bibliography{references}
\end{document}